\documentclass[11pt]{spie}

\def\simlt{\lower.5ex\hbox{$\; \buildrel < \over \sim \;$}}
\def\simgt{\lower.5ex\hbox{$\; \buildrel > \over \sim \;$}}
\newcommand{\citep}[1]{{\cite{#1}}}

\newcommand{\degree}{\ifmmode {^{\circ}} \else {$^{\circ}$} \fi}
\newcommand{\degrees}{\ifmmode {^{\circ}} \else {$^{\circ}$} \fi}

\newcommand{\unit}[1]{\ifmmode {\rm\ #1\,} \else {$\rm #1$} \fi}
\newcommand{\quarter}{\ifmmode {\frac{1}{4}} \else {$\frac{1}{4}$} \fi}
\newcommand{\arcmin}{\ifmmode {^\prime} \else {$^\prime$} \fi}
\newcommand{\arcsec}{\ifmmode {^{\prime\prime}} \else {$^{\prime\prime}$} \fi}

\newcommand{\tten}[1]{\ifmmode {\times 10^{#1}} \else {$\times 10^{#1}$} \fi}
\newcommand{\tentothe}[1]{\ifmmode {10^{#1}} \else {$10^{#1}$} \fi}

\newcommand{\doublet}{\ifmmode {\lambda\lambda} \else {$\lambda\lambda$} \fi}
\newcommand{\singlet}{\ifmmode {\lambda} \else {$\lambda$} \fi}

\usepackage{pslatex} \usepackage{graphicx} \usepackage{epsfig}
\usepackage{hyperref} \usepackage{epsf} \usepackage{ifpdf}

\title{A brief history of the search for extraterrestrial intelligence
and an appraisal of the future of this endeavor}

\author{Stuart Bowyer\\Professor of Astronomy (Emeritus)\\Space Sciences
Laboratory\\University of California\\Berkeley, CA 94720 USA}

\authorinfo{{\vspace{-0.3in}\\\ \\Email: {\em bowyer@ssl.berkeley.edu}\\
Telephone: +1 510 642-1648}}

\begin{document} \maketitle

The idea that  credible searches for Extra-Terrestrial Intelligence (ETI)
could be carried out were laid out in detail in the (now classic) paper
by Morrison and Cocconi (1959) \cite{Cocconi59}. They suggested using
the radio band for these searches. Since then radio searches have been
carried out by over sixty different groups. No signals from ETI’s have
been identified. Most searches did not have high sensitivity and it is not
surprising that  ETI signals were not detected.  It is important to note,
however, that these efforts were instrumental in developing new technical
capabilities and they helped generate wide interest in this field. In
this paper I will briefly discuss the more sensitive searches that have
been carried out and some of the other searches that are arguably quite
innovative or have been influential in some other manner.

\subsection*{Ohio State University}

In the early 1950’s John Kraus at Ohio State built a very crude radio
telescope for a search for extra terrestrial intelligence. However,
he realized that this telescope had severe limitations and consequently
he developed a new telescope for this search which became operational
in 1961. This telescope used  a very unique design which would not be
recognized as a telescope by most people. It used embedded metal mesh
laid on a flat ground area. A prime advantage of this design was its
ease of construction and hence its low cost. In the early 1960’s
Bob Dixon joined Kraus and was placed in charge of the data analysis
effort. Extensive observations were carried out and analyzed.

In 1977 a very odd signal was detected (dubbed the “Wow! space
signal”). There was no obvious terrestrial explanation for
this and it was widely reported in the popular press as an ETI
signal. Unfortunately the group lost its National Science Foundation
funding shortly thereafter. It was realized that the telescope had quite
limited sensitivity and it would need to be replaced for a significant
SETI program to be carried out. Lacking funding the facility was
closed. Perhaps the most significant contribution of this effort was
the publicity generated by the Wow! source and the raising of public
awareness of SETI searches.

\subsection*{Suitcase SETI/SENTINAL/META/BETA}

In the mid 1970’s Paul Horowitz
began a SETI project at Harvard University and in 1978 he carried out
a very limited search at ARECIBO. In 1981 he obtained funding from
NASA and the Planetary Society to build a high resolution spectrometer
which he dubbed Suitcase SETI.  In 1982  he installed this device at
ARECIBO  and searched for an ETI signal from 250 locations. In 1983
this instrument was mounted on  Harvard’s  26 meter telescope at the
University’s Oak Ridge Observatory and a search (dubbed SENTINAL)
was begun. Suitcase SETI and SENTINAL covered only 2 kHz bandwidth
which was a severe limitation. Hence a new spectrometer was developed
with a total of 8 million channels with 0.05 Hz resolution and 400 kHz
of instantaneous bandwidth. This device was dubbed META (Megachannel
ExtraTerrestrial Assay). Development of a spectrometer with even higher
resolution and a wider band-pass was begun in 1991 and was completed in
1995. This device, dubbed BETA (Billion-channel ExtraTerrestrial Assay),
was then mounted on the Oak Ridge telescope. Observations with this system
were carried out until 1999 when the telescope was blown over by strong
winds and was compromised. This ended the Harvard Group’s radio SETI
project. Perhaps the most significant result of their efforts was the
development of very high resolution spectroscopy for use in SETI.

\subsection*{SERENDIP}

At Berkeley I began a  SETI project (named SERENDIP, an acronym for
the Search for Extraterrestrial Radio Emissions from Nearby Developed
Intelligent Populations) in 1980. Our strategy was to use data obtained
in an observatory’s regularly scheduled astronomical programs. Data
acquired were analyzed off-line at the UC Berkeley Space Sciences
Laboratory. A commensal SETI program such as this is not free to choose
observing frequencies and sky coordinates. However, in view of the
plethora of postulated frequency regimes for interstellar communication
and the large number of potential sites for civilizations which have
been suggested, this is not necessarily a disadvantage.

The SERENDIP program began with a simple set of hardware with limited
capabilities. But it was a begun at a time when very few searches were
being attempted, and those that were being carried out were only done
intermittently. We were able to obtain substantial amounts of observing
time and we collected large quantities of data. Even more important was
our continuing work to develop more powerful instrumentation and signal
detection software. Our first instrument  used data from the University of
California Hat Creek Telescope. As our instrument capabilities progressed,
we were fortunate in obtaining time on  NRAO’s 300-foot telescope at
Green Bank West Virginia. We operated at this facility from 1986 until
it’s collapse due to a structural failure in 1988. (The tabloid press
attributed the collapse of this telescope to alien forces who resented
the fact that they were being tracked by our instrumentation).

The combination of our SERENDIP II instrumentation with this very large
telescope  resulted in a very high sensitivity search. In fact, our
searches have consistently been among the most sensitive in operation.
In Table \ref{tab1} (reprinted from Bowyer (2011)\cite{Bowyer11}),
I provide  a comparison of our search at Green Bank with other
instrumentation operating at that time.

In 1991 we installed an upgraded version of our instrument at the Arecibo
radio telescope in Puerto Rico. We have continued observations at this
facility where our latest and most sophisticated instrument continues
to  obtain data.

\begin{table}[htb] \begin{center} \caption{Summary of SETI programs,
circa 1986.\label{tab1}} \ifpdf \epsfig{file=bowyer_tab1.pdf} \else
\epsfig{file=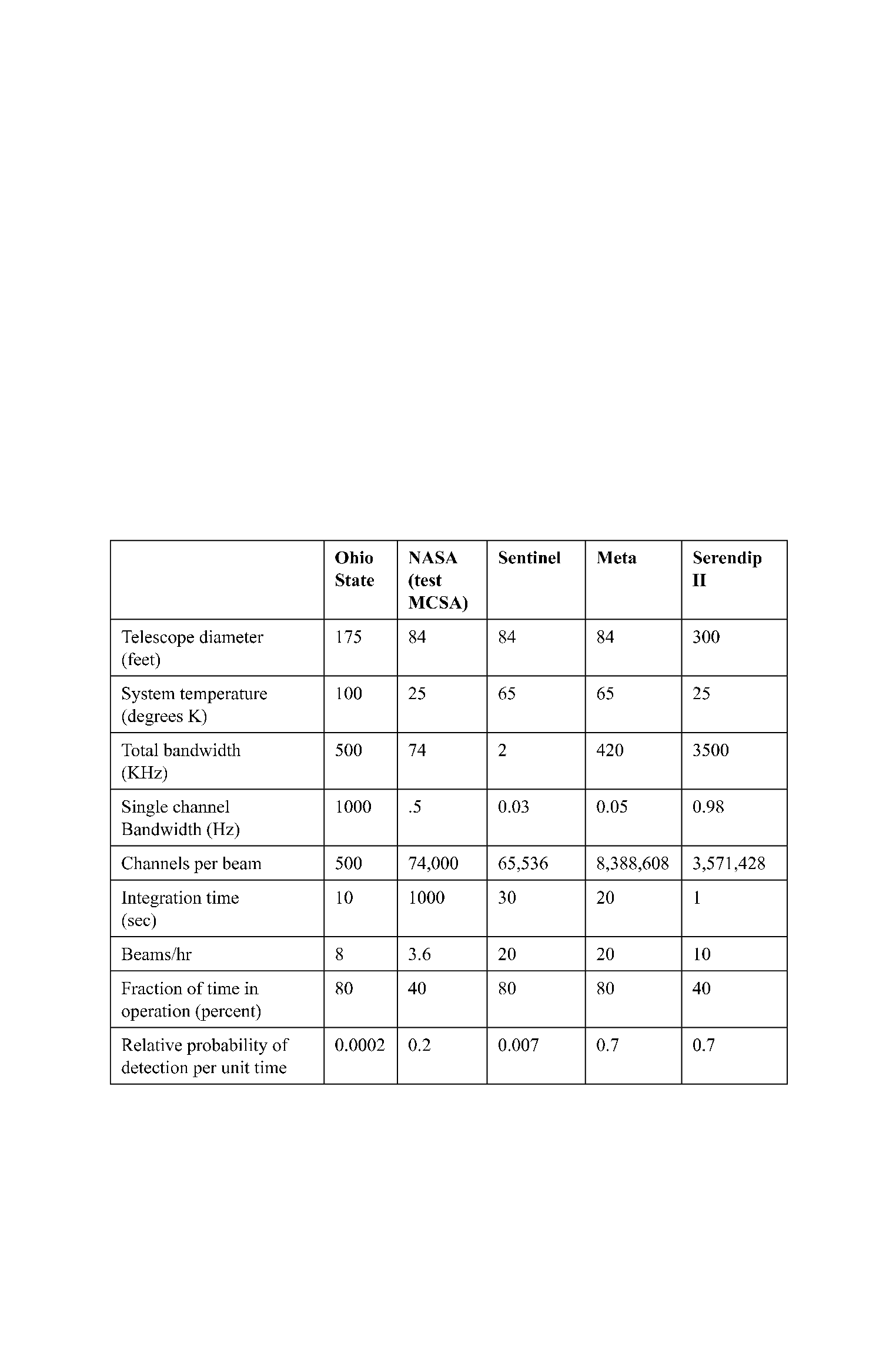} \fi \end{center} \end{table}

\subsection*{Bologna}

A group at the Institute of Radio Astronomy in Bologna, Italy led
by Stelio Montebugnoli has initiated a substantial SETI effort. In
this work they used the SERENDIP IV system which I arranged to have
transferred to them when we removed this instrument from the Arecibo
telescope for replacement with our improved SERENDIP V system. The
Bologna observations began in 1998 and are continuing today. They
are primarily carried out on a 32 meter dish in Medicina, Italy. This
group is pioneering efforts to use the the Karhunen-Loeve transform to
extract signals from noise. In principle the use of this transform has
substantial advantages over the the use of other types of transforms.
Unfortunately, this transform is extremely demanding computationally. A
straight-forward implementation of a system using this algorithm requires
an exponential increase in computing power as the quantity of the data
is increased. This is clearly untenable. The Bologna Group is having
some success in overcoming this problem.  A more detailed description
of this search is provided in Montebugnoli (2011)\cite{Montebugnoli11}.

\subsection*{NASA Ames}

Studies for a  SETI program to be carried out at the NASA Ames facility
began in the late 1970’s.  This program  was initiated at a low level
to examine various possible systems to carry out a SETI search. Another
extremely important part of this effort was the initiation of a program
for the development of data analysis systems.  The development of  the
instrumentation went through several different configurations as technical
problems were discovered in the various approaches that were investigated.
In 1991 development of the instrument was begun in earnest with funding of
millions of dollars per year.   The program went through a number of name
changes beginning when a group at NASA’s JPL’s facility joined the
effort, and again when NASA’s funding was withdrawn in 1993. Private
funding was then found to support this effort and the instrumentation
previously developed was placed on a number of different telescopes.
It is not entirely clear when actual observations (as opposed to test
trials) with this instrumentation began, but significant searches were
eventually carried out. A description of this search and its capabilities
when fully implemented and installed on the Allen telescope (described
below) are provided by Tarter (2011)\cite{Tarter11}.

\subsection*{ATA}

Paul Allan (the co-developer of Microsoft) funded the development of
what was envisioned to be the world’s largest SETI effort dubbed the
Allen Telescope Array. The facility was intended to consist of 350 radio
antennas to be located at the University of California’s Hat Creek Radio
Observatory. Because of a substantial number of technical difficulties
in the development of this array, Allan’s funding was exhausted after
only about twenty of these telescopes were developed to the point that
they were capable of being operated as an array. The ATA was officially
dedicated in late 2007 and test observations were begun. The data
analysis systems developed at NASA Ames were set up to use the output
from these observations. Tragically, in early 2011 NASA stopped funding
this program. New private funding was sought for this effort but none
could be found. Consequently this search had to be terminated.

\subsection*{Other types of searches}

Searching for ETI signals in other bands of the electromagnetic spectrum
has been proposed. Suggestions range from a search for pulsations in the
the optical flux in stars to the search for signals in the anti-neutrino
flux distribution. Of all these suggestions, the only actual non-radio
searches that have been carried out are searches for optical pulsations
in the output of stars. While arguments can be made for these optical
searches, it is my belief that these arguments are not strong. They
usually require that huge increases in laser emitting technology, far
beyond that envisioned in the most ambitious Star Wars laser programs,
will be developed.

An adjunct suggestion to the requirement that super mega-lasers  will
have to be developed is the hope that very large optical mirrors will be
feasible. One suggestion is that a parabolic mirror be built on the Moon
using a slowly rotating solution of molten salts with an ultra thin coat
of silver as the reflecting material.  I will not discuss these optical
projects in this paper.

\subsection*{A personal appraisal of the future of this endeavor}

By 2010 several radio searches were underway or were close to being
finalized which would have provided a huge increase in detection
capabilities. In addition, a major discovery of import was the detection
of a surprising number of planetary systems. This increases the potential
locations for extraterrestrial intelligence. However, the systems that
have been discovered show that the formation of planetary systems is
complex and results in many systems that are unstable. In a worst case
scenario for SETI, there will be far more stars with planetary systems
but few planets with stable orbits. An additional complexity is the
requirement, based on very general grounds, that water is a necessity
for the development of life. But the origin of water on Earth appears
to be the result of an odd set of circumstances which are not expected
to occur in virtually all planet formation scenarios. Despite these
uncertainties most workers in the field continue to be optimistic
regarding the posibility of eventually detecting an ETI signal.

Tragically, the search which was to be carried out with the Allen
Telescope at the Hat Creek observatory was canceled for funding
reasons. Other searches have also been canceled. Nonetheless, work
continues by groups that are developing new instrumentation and new
types of data analysis software. It is expected that these efforts will
ultimately result in substantially more sensitive searches.

Project Argus, a search effort coordinated by the non-profit SETI
League, encourages the development of SETI searches by amateurs. These
systems typically use small TV dishes. At this point the SETI League
coordinates the efforts of almost 150 operational searches throughout
the world. A fuller description of this effort is provided in Shuch
(2011)\cite{Shuch11}.

The Berkeley SERENDIP Program continues to acquire large amounts of
very sensitive data. Eric Korpela leads an effort (described in this
symposium) to develop new techniques for data analysis. This work has
been successful on several levels but the reduction and analysis of the
ongoing data set is progressing slowly.\cite{Korpela11}

Robert Dixon at the Ohio State Radio Observatory is developing a new type
of telescope for SETI. This is a high gain, omnidirectional array. It
is inherently less expensive than a dish and has no large or moving
parts. It has no tight mechanical tolerances and requires no mechanical
maintenance. The array itself is uniquely suited to a SETI search. The
challenge in using this type of array is that it  is extremely demanding
of computing power. Dixon currently has developed an array with twenty
four elements but there is no way to extract the signals from the data
stream using current computing capabilities. The group is investigating
various approaches which will reduce the computing demands of this
system to a reasonable level.  A more complete description of the work
is provided in Dixon (2011)\cite{Dixon11}.

In conclusion, it is possible that one of the currently operating searches
may detect an ETA signal. However,  given the intrinsic limitations of
these searches, this outcome seems unlikely. However, it is reasonable
to hope that in the long term (say in 50-100 years) instrumentation and
computing schemes (perhaps using some of the techniques discussed herein)
will have been developed to the point that large parts of the sky can
be simultaneously searched with high sensitivity and efficiency. At
that point one can be reasonably optimistic that a true ETI signal will
be discovered.

\bibliography{korpela} 
\bibliographystyle{spiebib}

\end{document}